\documentclass{aa}

\usepackage{graphicx}
\usepackage{color}
\usepackage{natbib}
\usepackage{amsmath}
\begin{document}
\title{Two-dimensional solar spectropolarimetry with the KIS/IAA Visible Imaging Polarimeter}

\author{C. Beck\inst{1,2}, L.R. Bellot Rubio\inst{3} , T.J. Kentischer\inst{4}, A. Tritschler\inst{5}\and J.C. del Toro Iniesta\inst{3}}
\institute{Instituto de Astrof\'isica de Canarias (IAC)         
           \and 
           Departamento de Astrof{\'i}sica, Universidad de La Laguna
           \and
           Instituto de Astrof\'isica de Andaluc\'ia (CSIC)          
           \and
           Kiepenheuer--Institut f\"ur Sonnenphysik
                     \and
           National Solar Observatory/Sacramento Peak
}
\titlerunning{Solar spectropolarimetry with the Visible Imaging Polarimeter}
\authorrunning{Beck et al.}
\offprints{C. Beck}

\date{\today}
%###############################################################################
%
%     ABSTRACT
%
%##############################################################################
\abstract {Spectropolarimetry at high spatial and spectral resolution
  is a basic tool to characterize the magnetic properties of
  the solar atmosphere.}  {We introduce the KIS/IAA Visible Imaging
  Polarimeter (VIP), a new post-focus instrument that upgrades the
  TESOS spectrometer at the German Vacuum Tower Telescope (VTT) into a
  full vector polarimeter. VIP is a collaboration between the
  Kiepenheuer Institut f\"ur Sonnenphysik (KIS) and the Instituto de
  Astrof\'isica de Andaluc\'ia (IAA-CSIC).}  {We describe the optical
  setup of VIP, the data acquisition procedure, and the calibration of
  the spectropolarimetric measurements. We show examples of data taken
  between 2005 and 2008 to illustrate the potential of the
  instrument.}  {VIP is capable of measuring the four Stokes profiles
  of spectral lines in the range from 420 to 700~nm with a spatial
  resolution better than 0\farcs5. Lines can be sampled at 40
  wavelength positions in 60 s, achieving a noise level of about $2
  \times 10^{-3}$ with exposure times of 300 ms and pixel sizes of
  $0\farcs17 \times 0\farcs17$ ($2\times2$ binning). The polarization
  modulation is stable over periods of a few days, ensuring high
  polarimetric accuracy. The excellent spectral resolution of TESOS
  allows the use of sophisticated data analysis techniques such as
  Stokes inversions. One of the first scientific results of VIP 
    presented here is that the ribbon-like magnetic structures of
  the network are associated with a distinct pattern of net
  circular polarization away from disk center.}  {VIP performs
  spectropolarimetric measurements of solar magnetic fields at a
  spatial resolution that is only slightly worse than that of the
  {\em Hinode\rm} spectropolarimeter, while providing a 2D field field of view
  and the possibility to observe up to four spectral regions
  sequentially with high cadence.  VIP can be used as a stand-alone
  instrument or in combination with other spectropolarimeters and
  imaging systems of the VTT for extended wavelength coverage.}

         
\maketitle
\section{Introduction}\label{sec:introduction}
Measurements of the solar magnetic field and the determination of
the thermal and kinematic properties of the magnetized atmosphere is a
persistently challenging goal. The combination of high spectral and
angular resolution is of vital importance for an accurate
characterization of the atmospheric parameters in different
structures, from the tiny magnetic elements of the quiet Sun to the
largest sunspots. However, only a few ground-based instruments are
capable of high-precision polarimetric observations with an angular
resolution of $0\farcs5$ or better.

Two different concepts have been followed to obtain measurements of
the full Stokes vector at high spectral re\-so\-lution: imaging and
grating (slit-based) spectropolarimeters. Modern imaging
spectropolarimeters employ single \citep[see, e.g.,][]{mickey+etal1996} or
multiple Fabry-P\'erot etalons (FPIs) for fast tuning and high
transmission. Several FPI spectropolarimeters are operated at present:
the G\"ottingen spectropolarimeter
\citep[GFPI,][]{gonzalez+kneer2008}, the Interferometric BIdimensional
Spectrometer \citep[IBIS,][]{cavallini2006,reardon+cavallini2008}, and
the CRisp Imaging Spectro-Polarimeter
\citep[CRISP;][]{scharmer+etal2008}. These instruments differ mostly
in how the FPIs are mounted (telecentric or collimated) and in the
spectral resolution.

In the category of slit-based spectropolarimeters, we have the
Advanced Stokes Polarimeter \citep[ASP,][]{skumanich+etal1997}
  which has been superseded by the the Diffraction-Limited
Spectro-Polarimeter \citep[DLSP,][]{sankarasubramanian+etal2003},
  the multiline polarimetric mode of THEMIS
  \citep[MTR,][]{rayrole+mein1993,lopezariste+etal2000}, the
POlarimetric LIttrow Spectrograph \citep[POLIS,][]{beck+etal2005b},
the Spectro-Polarimeter for Infrared and Optical Regions
\citep[SPINOR,][]{socas+etal2006}, and the Facility InfraRed
Spectropolarimeter \citep[FIRS,][]{jaeggli+etal2008}. DLSP and POLIS
work at a fixed wavelength in the visible (\@630\,nm), while SPINOR
and FIRS allow to observe different spectral regions in the visible
and the infrared. The Tenerife Infrared Polarimeter
\citep[TIP,][]{martinez+etal1999,collados+etal2007} can observe any
part of the near-infrared spectrum from 1 to 2.2 $\mu$m using the main
spectrograph of the German Vacuum Tower Telescope. It regularly
reaches the diffraction limit at these wavelengths (about 0\farcs6).

All ground-based instruments suffer from wavefront distortions caused
by turbulence in the Earths' atmosphere.  Adaptive optic systems
\citep[AO;][]{vdluehe+etal2003,scharmer+etal2003,rimmele2004a} are
mandatory to improve the image quality, but the compensation is only
partial and sub-arcsec resolution is hard to reach. Whereas in
slit-based spectropolarimetry no further improvement can be achieved
beyond what is delivered by the AO system, two-dimensional
spectropolarimetric observations can benefit from post-facto image
reconstruction techniques like de-stretching, speckle deconvolution
\citep[see, e.g.,][]{keller+vdl1992,mikurda+etal2006,puschmann+sailer2006,gonzalez+kneer2008}, and MOMFBD \citep[][]{vannoort+etal2005}. At present, the only
instrument capable of delivering continuous measurements at 0\farcs3
resolution is the spectropolarimeter (SP) of the {\em Hinode}
satellite \citep{kosugi+etal2007}. The Imaging Magnetograph eXperiment
\citep[IMaX,][]{jochum+etal2003,pillet+etal2004,herrero+etal2006}
aboard the SUNRISE balloon \citep{gandorfer+etal2006} recently
completed its first flight and took data of even higher
resolution thanks to its 1-m telescope.

However, accurate spectropolarimetry does not only depend on spatial
resolution, but also on the precision achieved in the determination of
the polarization state. This is typically quantified by the
root-mean-square (rms) noise of Stokes $QUV$ at continuum wavelengths
or by the signal-to-noise ratio (SNR) in the intensity spectrum.
  Slit spectropolarimeters have the advantage that the spectrum is
  detected instantly. Hence, an increase in exposure time or the
  accumulation of spectra do not compromise the spectral purity or
  integrity of the data.  With integration times of 20--30\,s, the rms
  noise can be reduced to a few 10$^{-4}$ of the continuum intensity
  \citep{bommier+molodij2002,khomenko+etal2003,marian+etal2008a,beck+rezaei2009}.
  For FPI-based instruments this is more difficult due to the
  sequential wavelength scanning. To obtain the SNR of a typical
  slit-spectrograph spectrum, a 2-D instrument has to use a similar
  integration time for {\em every} wavelength position, which usually
  makes the line scan unacceptably long ($>2$ min). Hence, for all
  2D-type spectropolarimeters a compromise has to be found between
  SNR, spectral sampling, and angular resolution. A SNR comparable to
  that of long-integrated ($\gg$ 5 sec) slit-spectra cannot be
  obtained with any currently existing 2D instrument.

However, these instruments also have great advantages compared to slit
systems. First, they are flexible in wavelength range. Even if the
various spectral regions can be observed sequentially only, the
cadence may be sufficiently high to treat them as simultaneous,
depending on the scientific problem. For slit-based systems, a change
of the spectral region usually requires changes in the setup, or is even
impossible because of fixed gratings (DLSP, POLIS, {\em
    Hinode}/SP). Second, 2D instruments provide spatially coherent
observations with much faster cadence than is possible with slit
systems. In high-resolution mode, when small step sizes are involved
(e.g., 0\farcs15 in the case of the {\em Hinode}/SP or the DLSP),
slit-spectrograph systems do not attain the cadence needed to
follow the solar evolution with sufficient spatial coverage. Fast
events, particularly in the solar chromosphere with time
  scales of 30\,s or less \citep{rutten+uitenbroek1991,carlsson+stein1997,wedemeyer+etal2004,
  woeger+etal2006, beck+etal2008}, can only be traced on very small
FOVs, whereas the spatial extent of the phenomenon may excess the scan
area easily.

The Triple Etalon SOlar Spectrometer
\citep[TESOS;][]{kentischer+etal1998, tritschler+etal2002} at 
the German VTT in Tenerife was a natural candidate for an upgrade into
a vector polarimeter. TESOS features high spectral resolution and is
the only triple-FPI system in operation to date. The unique
spectroscopic capabilities of TESOS have been used to study the
kinematic and thermal properties of several magnetic and non-magnetic
structures
\citep[see, e.g.,][]{schlichenmaier+schmidt1999,
schmidt+schlichenmaier2000,langhans+schmidt+tritschler2002,
schleicher+etal2003,tritschler+etal2004,schlichenmaier+bellot+tritschler2004,
bellotrubio+etal2006, mikurda+etal2006}. However, although valuable
for the analysis of photospheric velocity fields and temperatures,
these observations do not provide information about the magnetic
field.

Here we describe the KIS/IAA Visible Imaging Polarimeter (VIP), the
polarization package that converts TESOS into a full vector
spectropolarimeter. Section \ref{sec:motivation} gives an overview of
the optical layout of TESOS in polarimetric mode, focusing on the
modulation package and the detectors. The data acquisition and the
calibration procedure are explained in Sections \ref{sec:acquisition}
and \ref{sec:calibration}, respectively. In Sect.~\ref{sec:observations}, we present three examples of VIP observations to illustrate the capabilities of the instrument. Our conclusions are summarized in Sect.~\ref{sec:conclusions}.
\begin{figure*}
  \sidecaption
  \resizebox{12cm}{!}{\includegraphics{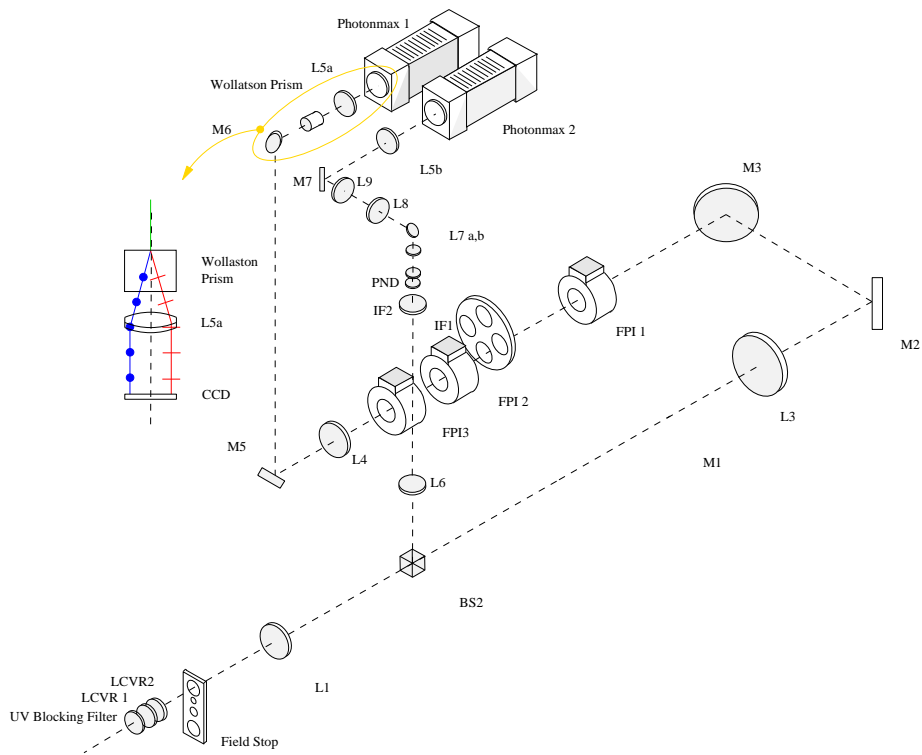}}
  \caption{Optical setup of TESOS in high resolution mode. The light
  undergoes four $45^\circ$ reflections at the mirrors M2, M3, M5 and
  M6. The Fabry-P\'erot etalons are indicated by the abbreviation
  FPI. The filter wheel (IF1) is located between FPI 1 and FPI2. The
  modulation package consists of two LCVRs near the telescope focal
  plane. PhotonMax 1 and 2 are the cameras of the narrow-band and
  broad-band channels, respectively. The Wollaston prism in front of
  PhotonMax 1 performs the polarization analysis.
\label{fig:tesos_3d}}
\end{figure*}
\section{Instrument Description\label{sec:motivation}}
TESOS is a Fabry-P\'erot spectrometer in a telecentric configuration
\citep{kentischer+etal1998, tritschler+etal2002}. Figure \ref{fig:tesos_3d} sketches the optical layout with its 
current components. The three etalons significantly reduce side-lobe
influence and permit the use of broad prefilters ($\sim$1\,nm FWHM)
with typical transmissions of 75\%.  In high resolution mode, the
spectral resolution of the instrument is of the order of 300.000
at \@632.8\,nm, comparable with classical slit spectrographs. Up to
four spectral lines can be observed sequentially thanks to the
motorized filter wheel (IF1 in Fig.~\ref{fig:tesos_3d}). Combined with
the Kiepenheuer Adaptive Optics System \citep{vdluehe+etal2003}, TESOS
is able to achieve a spatial resolution of about 0\farcs5 on a regular
basis.

Polarimetry was one of the science drivers of TESOS from the
beginning \citep[see][]{kentischer+etal1998}. To minimize instrumental
polarization, the four folding mirrors inside TESOS were arranged in
such a way that two sagittal 45$^{\circ}$ reflections are followed by
two tangential 45$^{\circ}$ reflections. This configuration does not
introduce spurious linear polarization signals and thus allows for
high precision polarimetry.

The KIS/IAA Visible Imaging Polarimeter (VIP) has been developed by
the Kiepenheuer Institut f\"ur Sonnenphysik and the Instituto de
Astrof\'{\i}sica de Andaluc\'{\i}a to upgrade TESOS into a full
vector spectropolarimeter. VIP consists of a modulation package,  
its electronics, a Wollaston prism, and the control software. With some
technical modifications, it could also be used at the main
spectrograph of the VTT.

\subsection{Modulation Package and Polarization Analysis}
The incoming light beam is modulated by two nematic Liquid Crystal
Variable Retarders (LCVRs) manufactured by Meadowlark. The LCVRs are
located directly in front of the field stop in a converging beam (f/64) 
close to the focal plane (LCVR1 and LCVR2 in Fig.~\ref{fig:tesos_3d}). 
A suitable blocking filter protects them from UV damage. The two 
LCVRs are mounted with their fast axes making an angle of 45$^\circ$.

LCVRs are electro-optical tunable retarders made of liquid crystal
molecules enclosed between two glass plates. Without an external
electric field, all molecules are aligned with their long axis
parallel to the glass substrate and maximum retardance is achieved. If
an external alternating electrical field is applied (square wave,
2\,kHz), the molecules {begin to tilt perpendicular to the glass substrate, which causes a reduction of the effective birefringence. The VIP LCVRs can be used in the range from 420 
to 700 nm; their retardance was measured as a function of the
applied voltage for two wavelengths (530\,nm and 630\,nm, see
Fig.~\ref{lc_ret_fig}).  The voltage needed for a specific retardance
and wavelength is then derived by a linear interpolation of the
calibration curves \citep{kentischer2005}.  At low voltages, the
interaction forces within the crystals are the dominant effects.
Therefore their relaxation time is longest in this regime (up to
60\,ms). To give the LCVRs enough time to reach the desired
retardance, the modulation sequence of VIP was chosen such that these
transitions occur when the instrument is busy with etalon settings,
camera readout, or disk writing.

Dual-beam polarimetry is achieved using a Wollaston prism before the
camera lens L5a and the detector (see Fig.~\ref{fig:tesos_3d}, zoom-in
area). It is oriented parallel to the fast axis of the first LCVR and
acts as a polarization analyzer.  The two orthogonal beams generated
by the Wollaston are imaged simultaneously onto the same detector. In
the data reduction process, they are combined to minimize
seeing-induced crosstalk \citep{lites1987}. To reduce image
distortions between the two beams, the Wollaston is made of calcite
instead of quartz.  The splitting angle of the prism, 1.37$^\circ$,
was optimized to fill the entire CCD.
\begin{figure}
   \centerline{\resizebox{8.8cm}{!}{\includegraphics{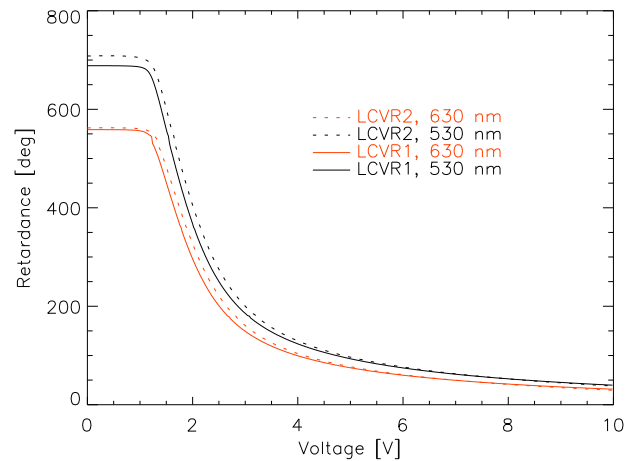}}}
  \caption{Calibration curves of the two LCVRs for the reference wavelengths of 530\,nm ({\em black}) and 630\,nm ({\em red}). The voltage was increased in steps of 0.02 V.\label{lc_ret_fig}}
\end{figure}
\subsection{Detectors\label{ssec:ccd}}
The PixelVision Pluto cameras used for TESOS were replaced in 2006.
The new detector system consists of two 16-bit, high frame rate
PhotonMax CCD cameras manufactured by Princeton Instruments/Acton.
Dark current is minimized by thermo-electrical cooling down to
$-80^{\circ}$\,C. The CCD image sensor is a backside illuminated,
512$\times$512 pixel array (e2v CCD97) with frame transfer technology
and a square pixel size of 16\,$\mu$m.  The wide dynamic range, large
full-well capacity (200\,ke$^{-}$ in traditional amplification mode)
and high quantum efficiency ($>$60\,\% in the wavelength range
400-850\,nm) of the cameras make them ideal for spectropolarimetry.

As a consequence of the different array and pixel size of the new
detector, the imaging optics in the narrow-band channel of TESOS 
had to be changed in order to allow for a FOV of about 40\arcsec\/ 
in diameter, while maintaining the pixel scale. The optimum
result was achieved with a $f = 169.7$\,mm lens, which provides
a pixel scale of 0\farcs086\,pixel$^{-1}$ and a FOV of 
$21\arcsec \times 41\arcsec$ in spectropolarimetric mode. 
\section{Spectropolarimetric Data Acquisition\label{sec:acquisition}}
To determine the Stokes vector, VIP takes four images $I_j$ ($j=
1\hdots4$) with the LCVRs in different modulation states. 
Following the IMaX strategy
\citep{pillet+etal2004}, the retardances of the first and second LCVR
are set to (315$^\circ$, 315$^\circ$, 225$^\circ$, 225$^\circ$) and
(305.264$^\circ$, 54.736$^\circ$, 125.264$^\circ$, 234.736$^\circ$),
respectively. This results in four linearly independent combinations
\begin{eqnarray}
\label{modulation}
\begin{pmatrix} I_1 \cr I_2 \cr I_3 \cr I_4 \cr \end{pmatrix} = 
\begin{pmatrix} 1 & 1/\sqrt{3} & 1/\sqrt{3}  & 1/\sqrt{3} \cr
1 & 1/\sqrt{3} & -1/\sqrt{3}  & -1/\sqrt{3} \cr 1 & -1/\sqrt{3} &
-1/\sqrt{3}  & 1/\sqrt{3} \cr 1 & -1/\sqrt{3} & 1/\sqrt{3}  &
-1/\sqrt{3} \cr
 \end{pmatrix} \cdot
\begin{pmatrix} I \cr Q \cr U \cr V \cr \end{pmatrix} \; ,
\end{eqnarray}
from which the Stokes $I$, $Q$, $U$ and $V$ parameters at the position
of the LCVRs can be derived.

The theoretical efficiencies are $(\epsilon_I, \epsilon_Q, \epsilon_U,
\epsilon_V) = (1, 1/\sqrt{3},1/\sqrt{3},1/\sqrt{3})$, with
$1/\sqrt{3}\sim0.577$. This provides the best modulation possible with
equal efficiency in all Stokes parameters
\citep{iniesta+collados2000}. The polarimeter can also be used to
measure only Stokes $I$ and $V$; the retardances of the LCVRs are then
($360^\circ$, $360^\circ$) and ($90^\circ$, $270^\circ$), respectively. To
improve the SNR, multiple images for each modulation state can be
accumulated by the control software. Up to now, we have used this
option only to obtain flat field data, since the extra read-out time
increases significantly the duration of the wavelength scans;
for observations, we use long exposures and the large full-well
capacity of the CCDs to reach a sufficiently high SNR.
\section{Data Calibration\label{sec:calibration}}
For the most part, the data reduction process is identical to the one
performed when the instrument is operated in spectroscopic mode, which
is described in detail elsewhere
\citep[see, e.g.,][]{tritschler+etal2004}. The reduction provides
flat-fielded and aligned images for each wavelength, modulation state,
and orthogonal beam. The alignment of the different scan steps is done
using the simultaneous images acquired with the broad-band
camera of TESOS (PhotonMax 2 in Fig.~1) as a reference. Here, we only
focus on the aspects that are special to spectropolarimetry.

The general approach for the polarimetric calibration of VIP is
similar to that of the other spectropolarimeters at the VTT (POLIS,
TIP, GFPI). The time-dependent instrumental polarization introduced by
the telescope and the remaining optics behind the exit window of the
telescope is corrected separately in two steps as described in detail
by \citet{beck+etal2005a}, and \citet{schliche+collados2002} or
\citet{beck+etal2005b}; for the GFPI see \citet{gonzalez+kneer2008}.

To determine the response function, $X$, of the polarimeter and the
optics behind the instrument calibration unit (ICU), 37 known
polarization states are created by the ICU and measured with VIP (see Fig.~\ref{calcurves}). The ICU is located right behind the exit
window of the evacuated telescope, and consists of a linear polarizer
and a zero-order quartz retarder. The retarder is rotated in 5$^\circ$
steps from 0$^{\circ}$ to 180$^\circ$, with the transmission axis of
the polarizer fixed along the terrestrial N-S direction.
Calibration curves for each of the LCVR states are obtained from the
intensity difference between the two orthogonal beams produced by the
Wollaston, normalized to their average value (see Fig.~\ref{calcurves}).
 A flat field correction is not needed for the calibration
measurements, because gain table variations affect the difference and
average images by exactly the same multiplicative factor, and thus they cancel out. Tests
with and without flat field correction yielded the same polarimetric
response down to our accuracy level. The subtraction of the dark
current is, however, crucial because it influences the relative
measurements of intensities in a non-linear way. The four calibration
curves can then be used to determine $X$ by a matrix inversion as
described in Appendix A.2 of \citet{beck+etal2005b}. Applied to the
four intensity measurements that result from subtracting the two
orthogonal beams for each LCVR state, $X$ gives the Stokes parameters
at the position of the ICU.
\begin{figure}
 \centerline{\resizebox{6.3cm}{!}{\includegraphics{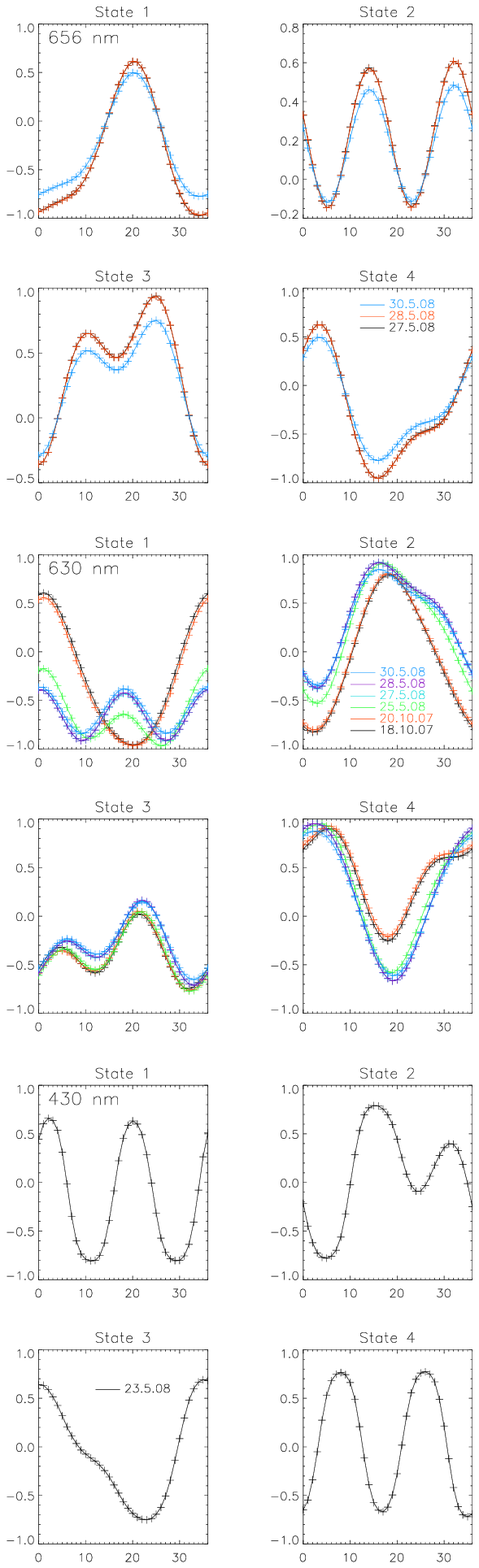}}} 
 \caption{Calibration curves for the four modulation states at 656,
  630, and 430\,nm (from {\em top to bottom}). Calibrations belonging to
  different days are given in different colors; in some cases the
  curves are nearly identical and cannot be distinguished from each
  other (e.g., \emph{black} and \emph{red} curve in \emph{top four
  panels}).\label{calcurves}}
\end{figure}

VIP differs from slit spectropolarimeters in its 2D FOV, which implies
that also the polarization properties can vary in both spatial
dimensions. We investigated the derived response functions for
subfields of the VIP FOV, but found no significant spatial variation
within the accuracy of $X$. Therefore we use the same response
function for every CCD pixel. We also found no significant trend with
wavelength inside the typical range used to scan a line, so we compute
$X$ only for the first wavelength position (usually a continuum
point).
\begin{figure}
  \centerline{\resizebox{8.4cm}{!}{\includegraphics{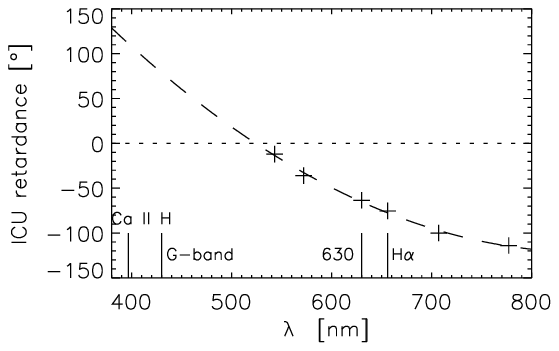}}}
  \caption{Retardance of the visible ICU at the VTT. \emph{Crosses} denote
           measurements, the \emph{dashed line} a parabolic fit.\label{ret_curve}}
\end{figure}

TESOS, and hence VIP, is prepared for sequential multi-wavelength
observations. This means that the calibration has to be done for as
many spectral regions as used in the observations, but a calibration
data set can be taken in less than 15 minutes. The variation of the
ICU retardance with wavelength is shown in Fig.~\ref{ret_curve}. The
parameters of a parabolic fit to the measured retardances are (688,
$-1.89537$, 0.00110836) for the offset, slope, and 2nd order
contribution. The chromatic wave plate of the ICU has zero retardance
around 520\,nm and cannot be used in this spectral region.

A critical issue for the accuracy of the measurements is the stability
of the polarization modulation over time. We consider here the nematic
LCVRs as the major source of variation, since the remainder of the
setup consists of ``stable'' massive objects like mirrors, the FPIs,
the Wollaston, or beam splitters (BS) that feed additional imaging
channels in front of TESOS. The LCVRs can change the polarimetric response of the instrument since their effective retardance is temperature dependent. The retardance decreases by about 1 degree for an increase in temperature by 1 K \citep[see, e.g.,][their Fig.~5]{heredero+etal2007}. Measurements at the ICU retarder showed a temperature increase of about 0.6$^\circ$ in two minutes on being exposed to sunlight before the temperature leveled off, as well as a reduction of retardance by 0.4$^\circ$ at 630\,nm comparing measurements in the early morning and at noon \citep{beck2004}. The light absorption of the LCVRs, and hence, their behavior, should be similar. During the observations, the LCVRs are permanently exposed to the sunlight, since no shutters or similar devices are located in front of them. There should thus only be some minor and fast heating effect before the start of the first observation in the early morning and a slight drift with time caused by the change of the light level during the day. 

The LVCRs of VIP are not located in a temperature-controlled housing, but the observing room as a whole is air-conditioned with a nominal temperature of 21$^\circ$. Temperature changes of the surrounding air  will presumably not exceed a level of one degree due to the air-condition. Taking one degree of temperature and thus one degree of retardance as the uppermost limit of variation, this translates into a deviation from the default modulation scheme by about three percent (e.g., $\cos 304^\circ/\cos 305^\circ$=0.975). This three percent is, however, not an absolute but a relative error; for instance for a polarization signal with an amplitude of 10\% of the continuum intensity, the resulting error would be only 0.3\%. Such an error level is comparable to the rms noise of the data (see Table \ref{tab:observations} below). The value of the rms noise actually implies that any short-term variations of temperature on the order of a few seconds cannot reach one degree, since it provides an upper limit for the contributions of {\em all} noise sources in addition to the thermal retardance effects. The main effect of the temperature changes will thus be a slow drift of retardance with time that presumably can be neglected for the polarimetric accuracy if the calibration is done right before or after the observations.

In order to verify the stability of the calibration, we collected calibration curves from several campaigns for wavelengths of 430\,nm, 630\,nm, and 656\,nm
(Fig.~\ref{calcurves}). The curves demonstrate that the calibration is
stable to within the measurement accuracy for up to two-three days
(e.g., 630 nm on Oct 18 and 20, 2007 or 656 nm on May 27 and 28,
2008). The large changes in the shape of the curves (compare 2007
(\emph{red}) and 2008 (\emph{blue}) in state 1, 630 nm) are due to the
additional BS used in 2008 to feed the imaging channels; the minor
changes between, e.g., May 25 and 30 were due to a re-adjustment of
the same BS. We conclude that daily calibration measurements should
suffice to maintain a high polarimetric accuracy for VIP if the
optical setup remains unchanged.

The measured polarization efficiencies are typically $(\epsilon_I,
\epsilon_Q, \epsilon_U, \epsilon_V) = (0.96, 0.53, 0.52, 0.56)$ at
630\,nm, slightly below the theoretical values but still very high.
The instrumental polarization caused by the VTT coelostat is
removed using the telescope model of \citet{beck+etal2005a}. If the
polarization level in a continuum window inside the observed spectral
range differs from zero, a correction for residual $I\rightarrow QUV$
crosstalk is applied by calculating the average $QUV/I$ ratio in the
continuum window and subtracting the corresponding fraction of the
intensity profile from $QUV$. This 
  crosstalk comes from two sources of calibration inaccuracies: the
  limitations of the geometrical telescope model and the high
  sensitivity of the first column of the response matrix to the
  retardance adopted for the ICU (see Fig.~4 of Beck et
  al.~2005b). The latter quickly leads to deviations of some percent
  (roughly 1\% of crosstalk per $1^{\circ}$ error in retardance). The
  determination of the ICU retardance can be improved with additional
  calibration data at different polarizer positions as described by
  \citet{beck+etal2005b}.
\begin{figure*}
\centerline{\resizebox{17.cm}{!}{\includegraphics{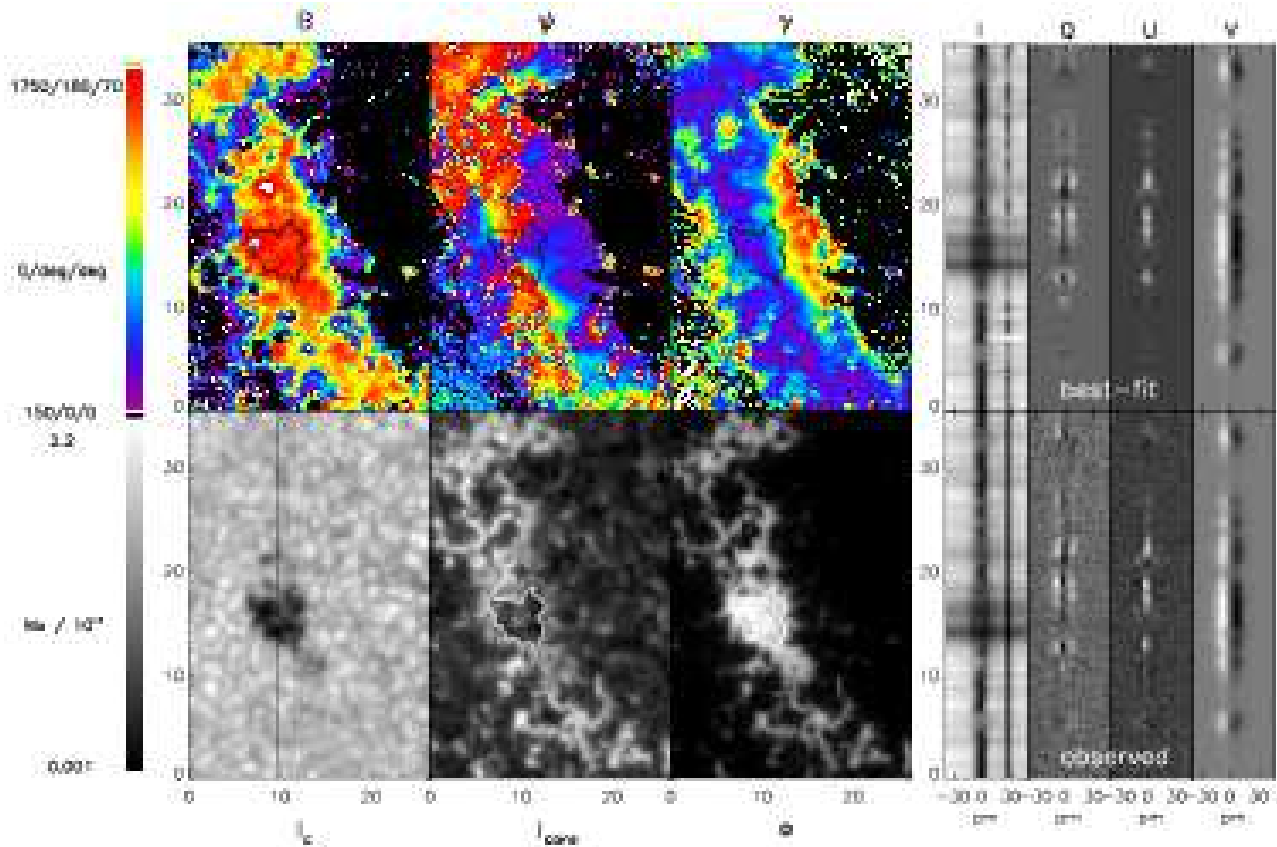}}}
\caption{\emph{Left panel}: Pore data taken in 2005.  \emph{Bottom
    row, left to right}: continuum intensity $I_{\rm c}$, line-core
  intensity of \ion{Fe}{i} 630.25\,nm, and longitudinal magnetic flux
  $\Phi$.  \emph{Top row, left to right}: field strength $B$, field
  azimuth $\psi$, and field inclination $\gamma$ in the LOS reference
  frame. The {\em black/white} contours outline the
    extension of the pore. The \emph{black vertical line} in $I_{\rm
    c}$ marks the cut whose spectra are shown in the \emph{right
    panels}. The observed and best-fit spectra are shown \emph{at
    bottom} and \emph{at top}, respectively.  \emph{Left to right}:
  Stokes $IQUV$. Spatial tick marks here and in all following figures
  are in arcsec.  \label{pore_fov}}
\end{figure*}
\section{Observations \label{sec:observations}} 
After commissioning, VIP has been used in several campaigns from 2005
until 2009. During the first test runs, VIP was a stand-alone
instrument devoted to the observation of the pair of \ion{Fe}{i} lines
at 630~nm. In later campaigns it has been operated simultaneously with
TIP and additional imaging channels as, for example, the G-band at
430.5\,nm and H$\alpha$ \citep[see, e.g.,][]{kucera+etal2008}. In the
following we describe and discuss three specific observations with VIP
to demonstrate its scientific potential. Details of the observations
are given in Table \ref{tab:observations}.

When coordinated with TIP, the spectrograph slit (of length
70\arcsec-80\arcsec) was oriented parallel to the long axis of the VIP
FOV, covering it completely. In the direction perpendicular to the
slit the coverage depends on the region scanned with TIP. The setup
for multi-instrument observations including VIP is described in
\citet{beck+etal2007a} and \citet{kucera+etal2008}.  

All observations are dark subtracted, flat fielded, and corrected for
transparency fluctuations during the scan. For the two data sets of
network no de-stretch prior to the polarization calibration has been
performed. Usually the observations are binned by a factor of two to
increase the signal-to-noise ratio, so the effective pixel size is
0\farcs17.
%*****************************************************************
% Table 1
%*****************************************************************
\begin{table}
\caption{Overview of the observations. I$_c$: continuum intensity.}
\begin{tabular}{l|lll}\hline\hline
Date             & Nov 3, 2005 & Oct 18, 2007& May 23, 2008 \\\hline
Target           & Pore        & Network     & Network \\
Position         & 10$^{\circ}$ & 20$^{\circ}$ & 30$^{\circ}$ \\        
VIP FOV          & 26\arcsec$\times$35\arcsec & 18\arcsec$\times$40\arcsec & 16\arcsec$\times$38\arcsec\\
Wavelength       & 630.2\,nm  & 630.25\,nm   & 430.34\,nm \\
Line(s)          & 2 \ion{Fe}{i}, O$_2$ &  \ion{Fe}{i}, O$_2$ & CH, \ion{Fe}{ii}   \\
Step Size        & 1.9\,pm     & 1.9\,pm      & 1.6\,pm  \\
Points           & 42+46          & 35           & 68 \\
Exposure         & 300\,ms     & 300\,ms      & 300\,ms \\
Cadence$^1$        & 72\,s+79\,s     &     70\,s           & 120\,s \\
Noise level      & 2$\cdot$10$^{-3}$\,I$_c$  & 2$\cdot$10$^{-3}$\,I$_c$   & 3$\cdot$10$^{-3}$\,I$_c$\\
TIP FOV          &  --         & 4.8\arcsec$\times$78\arcsec & 12.5\arcsec$\times$78\arcsec \\
Imaging         &   --          &  G-band & G-band \\\hline
\end{tabular}
$^1$Including disk write time; the scan time is about 10-20 s shorter, depending on the number of wavelength points
\label{tab:observations}
\end{table}
%****************************************************************
\subsection{Pore Observations\label{pore_sect}}
On November 3, 2005 we observed a pore (NOAA 10818) close to disk
center. VIP was used to scan the two \ion{Fe}{i} 630.2~nm lines
sequentially with a total of 88 wavelength points and a scan
time of about 65 s for each line. For the following, we only considered the line at 630.25\thinspace nm. In order to determine reliable magnetic parameters and to assess the performance of VIP, the Stokes $I$, $Q$, $U$ and $V$ profiles were inverted using the SIR code. We performed a 1-component
inversion with a variable stray light contribution $\alpha$ fitted by
the algorithm.  Depending on whether the polarization signal exceeded
a threshold of 0.85\,\% of the continuum intensity or not, either a
single magnetic or a field-free atmosphere was assumed.
\begin{figure*}
\centerline{\resizebox{14cm}{!}{\includegraphics{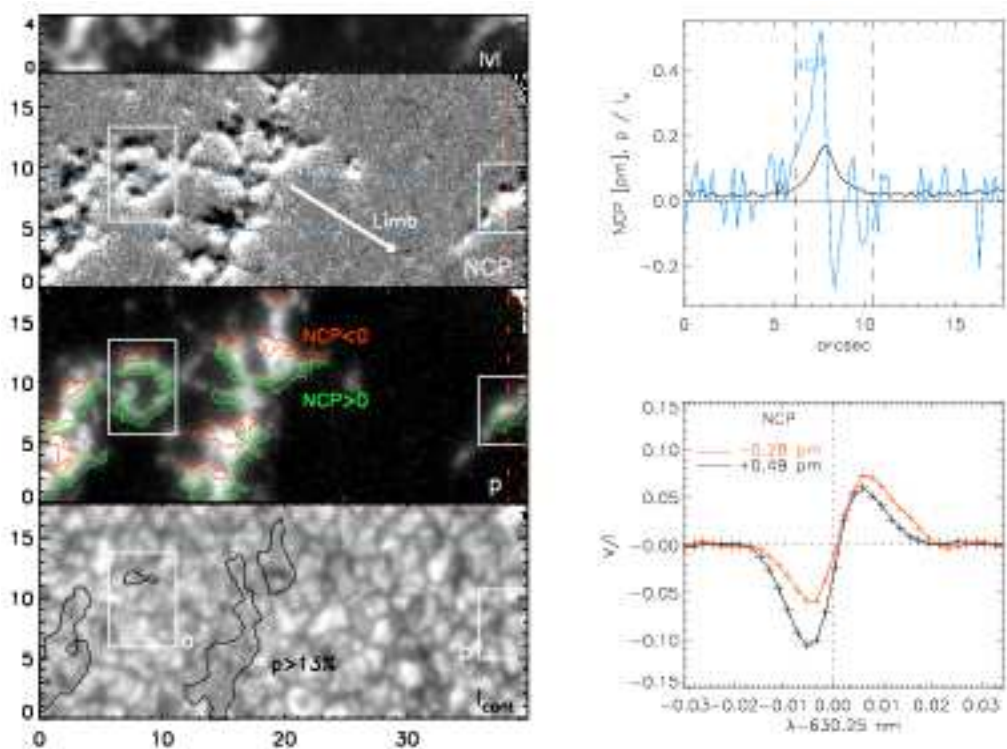}}}
\caption{\emph{Left panel}: Network data taken in 2007.  \emph{Bottom
    to top}: $I_{\rm c}$, polarization degree $p$, NCP, $|V|$ from
  TIP. {\em Red} and {\em green contours} in $p$ outline negative and
  positive NCP, respectively; \emph{black contours} in $I_{\rm c}$ a
  polarization degree above 13\%. \emph{The blue horizontal dashed
    lines} in NCP represent the TIP FOV. The direction to the
    limb is indicated by the {\em white arrow}.
  \emph{Right top}: NCP (\emph{blue}) and $p$ (\emph{black}) along the
  cut marked by a {\em vertical red dashed line} crossing the
    \emph{right white rectangle}, b. The {\em black vertical
      dashed lines} denote the lower and upper boundary of rectangle
    b. \emph{Right bottom}: Stokes $V$ profiles with maximum positive
  and negative NCPs in the cut. \label{network_fig}}
\end{figure*}

Figure \ref{pore_fov} displays the observed FOV (\emph{bottom panels})
and some of the inversion results (\emph{top panels}). The continuum
intensity (\emph{bottom left}) shows clear signs of abnormal
granulation surrounding the pore. The line-core intensity image
(\emph{bottom middle}) reveals brightenings that closely follow the
course of the intergranular lanes throughout most of the FOV. The
spatial resolution is sufficient to distinguish the ribbon-like
appearance of these structures, discovered by \citet{berger+etal2004a}
in G-band filtergrams taken at the Swedish 1-m Solar Telescope. A
comparison with the unsigned magnetic flux $\Phi = |(1-\alpha)\cdot B
\cdot\cos \gamma |$ (\emph{bottom right}) demonstrates that all the
strong line-core brightenings are related to the presence of magnetic
fields. This does not immediately imply that the upper layers of magnetic
elements are hotter than their surroundings, as not only the temperature but
also the Zeeman splitting or the shift of the optical depth scale in the presence of magnetic fields increase the line-core intensity. The flux concentrations reside predominantly in the intergranular
lanes. The field strength (\emph{top left}) ranges from 50\,G to about
2\,kG. The field azimuth (\emph{top middle}) and field inclination
(\emph{top right}) indicate radially oriented magnetic fields that
extend well beyond the visible border of the pore ({\em white/black} contours). The inclination
increases with radial distance from the center of the pore, reaching a
maximum of about 70$^\circ$ to the LOS.  As the pore was located close
to disk center, the corresponding orientation in the local reference
frame (i.e., with respect to the local surface normal) will not differ
much from the LOS values. In the \emph{right half} of
Fig.~\ref{pore_fov}, the observed and best-fit spectra along a cut
through the pore are shown for comparison.

We conclude that because of the fine spectral sampling and the high
SNR achieved, the spectra can be well analyzed with the SIR code,
allowing for an accurate determination of the magnetic field
parameters at high spatial resolution. A preliminary investigation of the effect of the transmission profile of TESOS on the observed spectra, mainly using the telluric oxygen line, suggested systematic variations of line widths or line positions across the FOV of 0.1\,--\,0.2\,pm \citep[see also][]{vdluehe+kentischer2000,tritschler+etal2002, pillet+etal2004,scharmer2006,puschmann+etal2006,reardon+cavallini2008}. This translates into an error in the field strength (velocity) of 50\,G (50 ms$^{-1}$) at 630\,nm, which is comparable to the error in the determination of the velocity or the magnetic field strength \citep[see, e.g.,][]{beck2006} and far below their intrinsic variation across the solar surface. A more detailed discussion of the instrumental effects on the retrieved physical parameters is postponed to a future publication.
\subsection{Network Observations}
The network observations taken on October 18, 2007 at about 10:00\,UT
were coordinated with TIP. A broad-band G-band speckle channel was
added in front of TESOS to provide context information with high spatial resolution. TIP was set to repeatedly scan a small region of
$\sim$5\arcsec\/ near the centre of the VIP FOV\footnote{See the TIP
archive for an overview of the TIP data (KIS home
page$\rightarrow$Observatories$\rightarrow$Data archives).}.

Figure \ref{network_fig} summarizes the observations: the continuum
intensity $I_{\rm c}$ (\emph{bottom left}), the polarization degree
$p$ (\emph{middle left}), and the net circular polarization (NCP;
\emph{top left}) in the \ion{Fe}{i} line at 630.25~nm. The NCP was
computed as the integral of Stokes $V$ over wavelength, multiplied by
the polarity of the field. The polarization degree was determined
  as
\begin{equation}
p = {\rm max} \left(\sqrt{Q^2(\lambda)+U^2(\lambda)+V^2(\lambda)}\right) / I_{\rm QS},
\end{equation}
with $I_{\rm QS}$ representing the quiet Sun continuum intensity. 
The small map at the \emph{very top left} shows the integrated $V$ 
signal of one TIP scan. Its location inside the VIP FOV is 
indicated by the \emph{blue horizontal dashed lines} in the NCP map.

 The NCP map reveals a distinct pattern, varying between
  positive and negative values on small spatial scales (e.g., {\em
    left white rectangle}, a). Negative NCP ({\em black}) appears
  preferentially toward the disk center and positive NCP ({\em white})
  toward the limb. The polarization degree map shows that the highest
  NCP values are not cospatial with the highest polarization degree,
  but rather flank it.  The effect is visualized in the {\em right
    panel} that shows NCP and $p$ values on a cut along the y-axis
  through the {\em right white rectangle}, b. The NCP reduces to zero
  and changes sign across the maximum of the polarization degree. The
  Stokes $V$ profiles corresponding to maximal positive and negative
  NCP in the cut are shown at the \emph{bottom right}; the pronounced
  asymmetry of the red and blue $V$ lobe is clearly visible in the two
  cases.  The spectral sampling is indicated by the crosses. 

  Figure \ref{ncp_sketch} offers a tentative explanation of how
  the peculiar NCP pattern could possibly be created. In the case
    of a flux sheet that is parallel to the limb, the LOS crosses
  both the canopy and the center of the sheet ({\em top row}). If the
  flux concentration is perpendicular to the limb ({\em bottom
    row}), the LOS passes through either the canopy {\em or} the
  central axis. This scenario provides fairly different gradients of
  magnetic field strength along the LOS. In addition, velocity
    gradients will presumably be encountered by the LOS at the
    boundary of the flux concentration; the field-free convective
    region below the canopy and the magnetic volume are expected to
    harbor upflows and downflows, respectively.  With these velocity
  gradients, the two ne\-ce\-ssary conditions for generating a
  NCP are present
  \citep[see, e.g.,][]{auer+heasley1978,almeida+lites1992,landolfi+landi1996}.
  The scenario, however, will have to be investigated in more detail
  before it can be taken for plausible. The detection of the variation
  of the NCP across magnetic flux concentrations has been
    possible thanks to the higher spatial resolution of VIP \citep[or
  {\em Hinode};][]{rezaei+etal2007a} compared to older data where the
  resolution was not sufficient to separate the canopy from the
  central axis.  \citep[see, e.g.,][]{sigwarth2001}. Interestingly, a
    variation of the Stokes $V$ area asymmetry across magnetic
    elements was inferred from the inversion of ASP measurements at 1\arcsec\/
    \citep{bellotrubio+etal2000}.
%
%*****************************************************************
% Figure 6
%*****************************************************************
\begin{figure}
  \centerline{\resizebox{5cm}{!}{\includegraphics{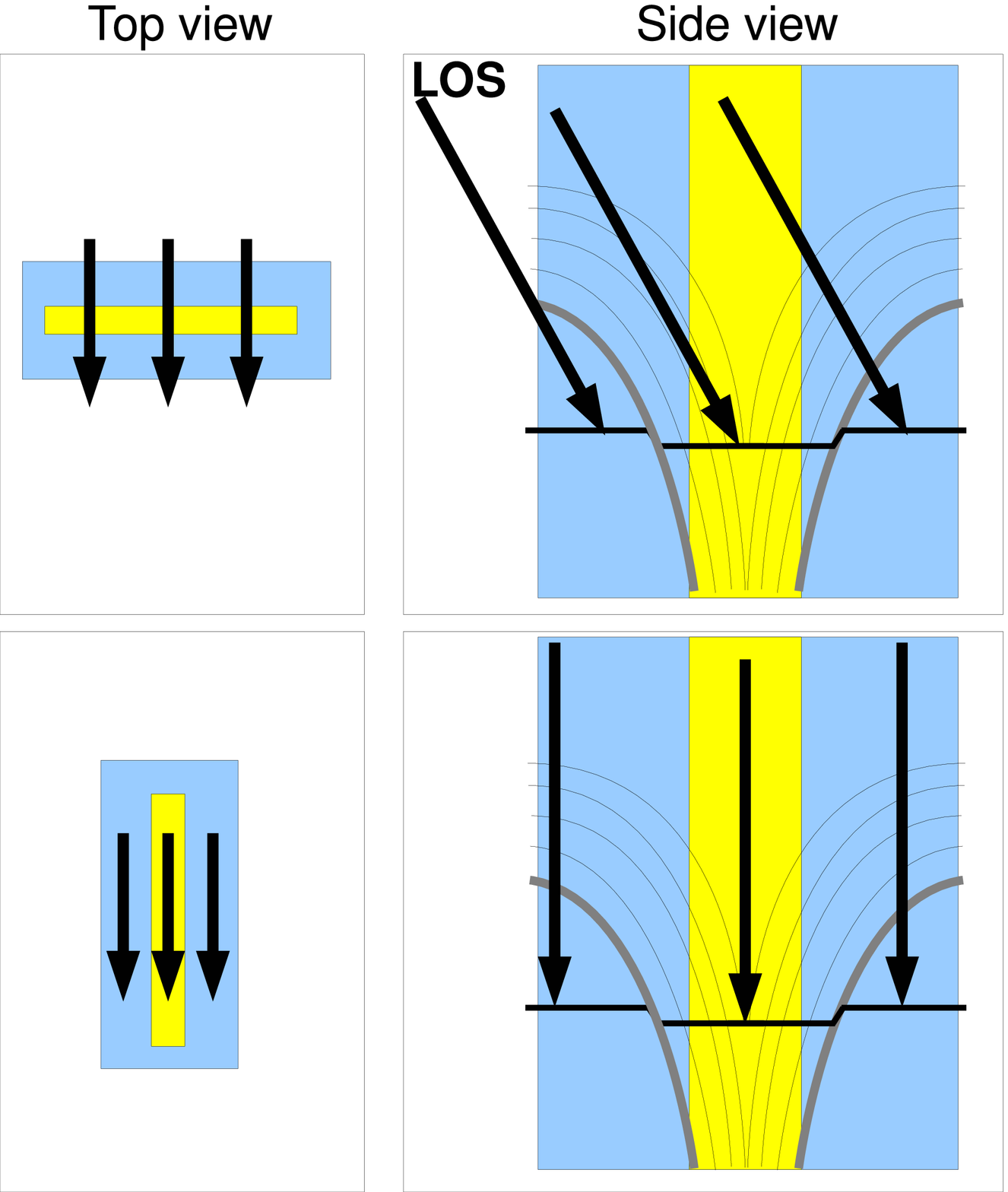}}}
  \caption{Dependence of NCP behavior on the orientation of
      magnetic flux sheets.  For an orientation of the sheet parallel to the limb ({\em top row}), the inclined LOS passes through
      both the central part ({\em yellow}) and the canopy ({\em
        blue}), for an orientation perpendicular to the limb
      ({\em bottom row}), it passes through either the canopy {\em or} the central axis.
      The LOS is inclined out of the paper plane in all but the upper
      right panel. \label{ncp_sketch}}
\end{figure}
%*****************************************************************
%*****************************************************************
\subsection{Network Observations in the G band}

VIP was used for spectropolarimetry in the G band on May 23, 2008. The
wavelength range covered several CH lines and one strong atomic
line near 430.32\,nm. The short wavelength of the observations led to
a rather low average intensity of about 490 counts, but the rms noise
in the polarization signal stayed at about 0.3\,\%. Again, VIP was
operated in coordination with TIP and an external speckle G-band
channel\footnote{In this observation, the spectral resolution
    of TESOS was lower than usual because of a mistake in the setup
    (the parallelism of the FPIs was not adjusted).}.
\begin{figure*}
\centerline{\resizebox{17.6cm}{!}{\includegraphics{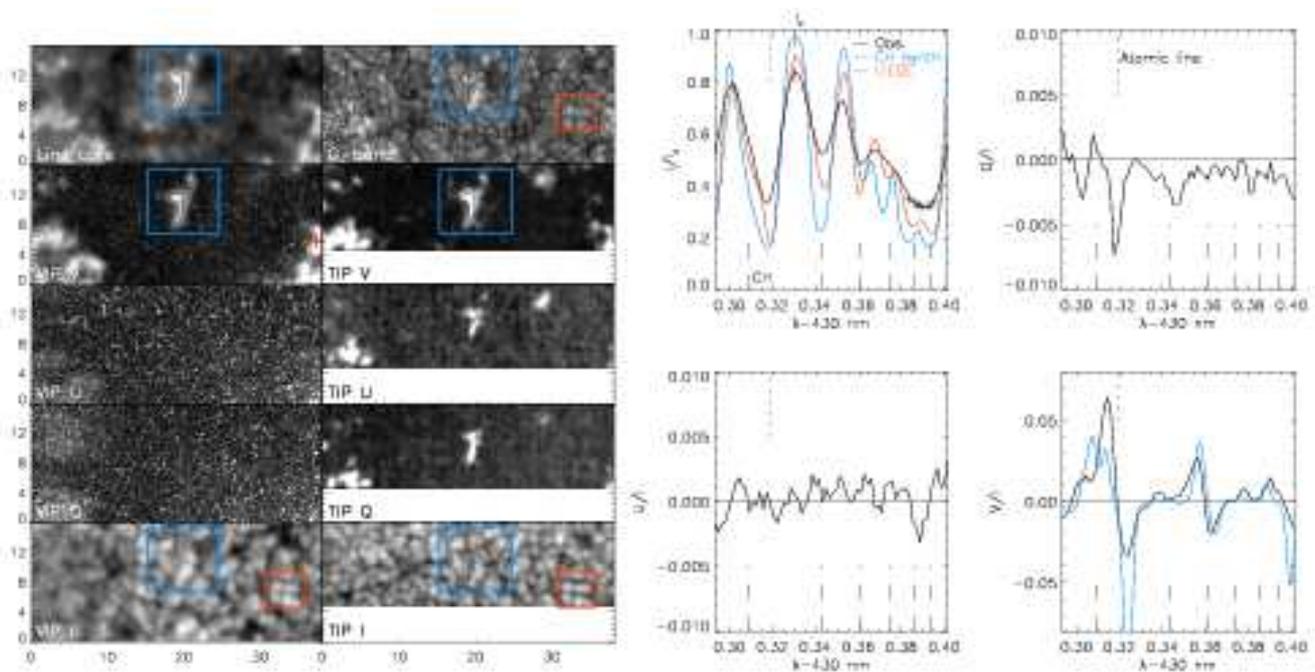}}}
\caption{{\em Left panel}: 2008 G-band observations.  {\em Left
    column, bottom to top}: Stokes $I$ and wavelength-integrated
  absolute $QUV$, and line-core intensity near 430.39 nm, as measured
  by VIP. {\em Right column, bottom to top}: Stokes $I$ and
  wavelength-integrated absolute $QUV$ signals from TIP, and
  speckle-reconstructed broad-band G-band image. The {\em blue} and
  {\em red rectangles} identify the features used for the data
  alignment. The {\em black contours} outline large $V$ signals in
  TIP. The {\em red cross} in the circular polarization map from VIP
  near (40$^{\prime\prime}$, 7$^{\prime\prime}$) marks the location of
  the Stokes $IQUV$ spectra shown in the {\em right panel}. {\em
    Black}: profiles observed with VIP; {\em red}: Li\`ege atlas
  profiles; {\em blue}: G-band synthesis (courtesy of A.~Asensio
  Ramos, IAC). CH lines are marked with {\em dashed vertical lines}
  from bottom up, one atomic line with a {\em vertical dotted line}
  from top down. In Stokes $I$, a short bar at 430.33 nm denotes the
  continuum wavelength, and the {\em black crosses} around
    430.39\,nm the line-core intensity wavelength range.
  \label{gband_fov}}
\end{figure*}

Figure \ref{gband_fov} shows an overview of the VIP and TIP
observations of a network region located at
$(x,y)=(-446^{\prime\prime}, -193^{\prime\prime}$), i.e.,
$30^\circ$ away from the disk center. In this case, the two instruments
covered a similar FOV. The ``continuum'' intensity for VIP was taken
at 430.33 nm, where the intensity in the observed wavelength range is
maximum. VIP did not reveal clear Stokes $Q$ and $U$ signals above the noise
level, in contrast to TIP. This may partly be attributed to the different
type of lines observed, the different spectral regions, or the
different SNRs achieved. The $V$ signals of both instruments, however, match closely. 

The Stokes profiles recorded by VIP can be used for quantitative
analyses even at the low light level prevailing in the blue. The
{\em right panel} of Fig.~\ref{gband_fov} shows sample spectra from a
relatively strong flux concentration marked with a {\em red cross} in
the Stokes $V$ map of VIP. Besides studying how the depth of the
various CH lines in Stokes $I$ depends on the physical properties of
the plasma and the magnetic field, the observed Stokes $V$ profiles
({\em bottom right}) can be used to verify theoretical models of CH
line formation. The overplotted {\em blue line} comes from a numerical
calculation by \citet{uitenbroek+etal2004}, for a height-independent
field strength of 1 kG. The $V$ signal was scaled down by a factor of
1.5 to better reproduce the observed CH lines redward of 430.36\,nm.
The large mismatch between observed and synthetic profiles
around 430.31\,nm is presumably caused by the parameters used to synthesize the atomic line of \ion{Fe}{ii} (A.~Asensio Ramos, private communication).
\section{Summary}\label{sec:conclusions}
The KIS/IAA Visible Imaging Polarimeter (VIP) is a new instrument for
2D spectropolarimetry of the solar atmosphere. It is used with TESOS,
the triple etalon spectrometer installed at the German Vacuum Tower
Telescope.  The polarimeter is based on a pair of nematic liquid
crystal retarders and a Wollaston prism. In combination with the
adaptive optics system of the telescope, VIP and TESOS provide full
Stokes vector measurements of spectral lines in the visible at a
resolution of 0\farcs5 or better. Using exposure times of 300~ms for
each modulation state and $2 \times 2$ binning, the noise level is
about 0.2\% of the continuum intensity. The four Stokes profiles can
be measured at 40 wavelength positions in about 60~s. The response
function of the polarimeter is determined using the instrument
calibration unit of the telescope, and turns out to be stable over a
few days.

The high resolving power and excellent performance of TESOS and VIP
make it possible to derive the magnetic field geometry and
obtain information about the atmospheric conditions by means of
inversion techniques. The Stokes spectra recorded by VIP show strong
asymmetries, as expected for high-resolution measurements. Thus, it
should be possible to determine vertical gradients of the atmospheric
parameters from them. Photospheric and chromospheric lines (e.g.,
H$\alpha$) can be observed sequentially thanks to the motorized filter
wheel of TESOS.

Usually VIP is operated in coordination with the Tenerife Infrared
Polarimeter and speckle imaging systems at the German Vacuum Tower
Telescope, providing multi-wavelength observations of the same
structures and processes. VIP could also be used for coordinated
observations with space-borne instruments such as {\em Hinode} or the
upcoming Solar Dynamic Observatory. Because of these reasons, 
we anticipate that future observations with VIP will be used to 
address several open questions in solar physics. 
\acknowledgements
The VTT is operated by the Kiepenheuer--Institut f\"ur Sonnenphysik,
Freiburg, Germany, at the Spanish Observatorio del Teide of the
Instituto de Astrof\'{\i}sica de Canarias. This work has been
partially funded by the Spanish Ministerio de Ciencia e Innovaci\'on
through project ESP2006-13030-C06-02 (including European FEDER funds) and through project AYA 2007-63881.

\bibliographystyle{aa}
\bibliography{references_vip}
\end{document}